# Discontinuous transition to shear flow turbulence

Bowen Yang[1,3], Yi Zhuang[1,3], Gökhan Yalnız[1], Vasudevan Mukund[1], Elena Marensi[2] & Björn Hof[1]

Depending on the type of flow, the transition to turbulence can take one of two forms: either turbulence arises from a sequence of instabilities or from the spatial proliferation of transiently chaotic domains, a process analogous to directed percolation. The former scenario is commonly referred to as a supercritical transition and frequently encountered in flows destabilized by body forces, whereas the latter subcritical transition is common in shear flows. Both cases are inherently continuous in a sense that the transformation from ordered laminar to fully turbulent fluid motion is only accomplished gradually with flow speed. Here we show that these established transition types do not account for the more general setting of shear flows subject to body forces. The combination of the two continuous scenarios leads to the attenuation of spatial coupling; with increasing forcing amplitude, the transition becomes increasingly sharp and eventually discontinuous. We argue that the suppression of laminar–turbulent coexistence and the approach towards a discontinuous phase transition potentially apply to a broad range of situations including flows subject to, for example, buoyancy, centrifugal or electromagnetic forces.

One of the earliest models for the onset of turbulence goes back to Landau[1] who envisioned a sequence of instabilities that gives rise to a large number of incommensurate modes and a stepwise increase in the flow's complexity. This view was later challenged and refined by Ruelle and Takens[2] who showed that chaotic dynamics already arises after a small number of bifurcations. Such universal routes[3] to chaos are commonly observed in flows dominated by body forces[4–6] (for example, buoyancy or centrifugal forces) that render the flow linearly unstable. The state of the flow is, in this case, described by an amplitude (for example, amplitude of convection rolls), which is a continuous function of the control parameter, provided that the underlying bifurcations are supercritical and this scenario is, therefore, generally referred to as a supercritical transition.

Conversely, in shear flows including those in pipes, channels and boundary layers, the transition can occur in the absence of a linear instability of the laminar flow. It arises 'subcritically'[7,8], or more precisely, turbulence develops from nonlinear, unstable solutions that are unrelated and dynamically disconnected from the laminar base flow[9–13]. As a consequence, local flow amplitudes scale discontinuously at the transition point. Despite this finite amplitude gap, the transition is nevertheless continuous: starting from fully turbulent flow, turbulence does not disappear abruptly with a decreasing Reynolds number ($Re = UD/\nu$, where for pipe flow $U$ is chosen as the bulk velocity, $D$ is the pipe diameter and $\nu$ is the fluid's kinematic viscosity). Instead, turbulence ceases to be space filling and the fraction of the flow that is turbulent decreases gradually, until it eventually reaches zero at the transition threshold. Hence, in these cases, the laminar–turbulent coexistence (LTC) renders this transition continuous. Moreover the spatiotemporal nature of the dynamics turns this problem into a phase transition (typically directed percolation[14,15]), which is described by an order parameter, that is, the turbulent fraction (TF). Conversely, in the absence of LTC, a subcritical transition is intrinsically discontinuous.

LTC, in turn, arises from spatial coupling and energy transfer between the laminar and turbulent regions. In pipe flow, turbulent structures of limited size—puffs—prevail by extracting energy from the upstream laminar flow[16,17]. The high-speed laminar fluid close to the centreline impinges onto the adjacent puff and gives rise to a peak in turbulent production at the puff's upstream interface[18]. A local manipulation of the upstream laminar velocity profile, adjusting its shape to resemble that of turbulence, directly intercepts the energy input and causes the puff to decay[18]. In particular, the energy advection at laminar–turbulent interfaces and the downstream profile recovery is a central aspect of localized structures in transition models[19–22].

[1]Institute of Science and Technology Austria (ISTA), Klosterneuburg, Austria. [2]The University of Sheffield, Sheffield, UK. [3]These authors contributed equally: Bowen Yang, Yi Zhuang. e-mail: bhof@ist.ac.at

Past investigations often either focused on supercritical scenarios, typically driven by body forces, or on the subcritical transition scenario via LTC. Although the distinction between the two transition scenarios may appear clear, most practical situations are more complicated; in addition to shear, flows are equally subject to body forces. Examples range from heating (cooling) pipes and flows around bends and corners, all the way to magnetohydrodynamic (MHD) flows in geo- and astrophysics. At the same time, studies of transition in simpler shear flows subject to such body forces reported the persistence of LTC[23–25] in the subcritical regime, inferring the robustness of the continuous scenario. However, as we will show in the following, standard body forces—irrespective if we consider buoyancy, centrifugal or electromagnetic forces—fundamentally alter the nature of the transition. Even though these different forces seemingly have nothing in common and alter the velocity profile in entirely different ways, surprisingly, the effect on transition is identical. Rooted in the suppression of energy flux, LTC, the hallmark of the shear flow transition, is marginalized, giving way to a much simpler discontinuous scenario.

We start our investigation with curved and heated pipes, two examples for which a linear instability of the laminar flow will eventually be encountered at large body force amplitudes. Our focus is solely on the impact of body forces on the subcritical transition scenario and, hence, on force amplitudes below the linear stability threshold. In the case of curved pipes, the centrifugal force that arises is known to substantially delay the subcritical transition via puffs and slugs[26], and past experiments have used pipe curvature to relaminarize turbulent flow[27]. More recently, direct numerical simulations[28] have shown that curvature not only delays the transition to puffs but it equally suppresses turbulent production at the laminar–turbulent interfaces and reduces the Reynolds number regime across which puffs are observable. Despite this suppression for the parameters investigated in this numerical study, puffs were reported, and hence, the coexistence of laminar and turbulent flows persisted. We performed laboratory experiments for a curvature of $R/A = 0.0173$ (where $R$ is the pipe radius and $A$ is the coiling radius), a value greater than that of the aforementioned simulations ($R/A = 0.01$), yet still below the threshold of linear stability[23,27–29] ($R/A \approx 0.025$). More precisely, our measurements were conducted in a helical pipe, composed of a semiflexible tube that is coiled around a cylinder of radius $A = 22.5$ cm, whereas the pipe has an inner radius of $R = 4$ mm. The helical tube had a pitch of 12 mm or $3R$ per coil (Extended Data Fig. 1). The tube has a length of $L = 98$ m and a dimensionless length of $L/R = 24{,}500$. The addition of flow visualization particles allows the discrimination between laminar and turbulent regions, and velocity fields were measured using particle image velocimetry. Further details are provided in the Methods. In the present case, the linearly stable laminar flow was perturbed directly at the inlet: before entering the curved section, the fluid passes through a 300-$R$-long straight tube with an identical diameter. At the beginning of the straight pipe, a small pin protrudes into the tube and assures fully turbulent inflow conditions provided that $\mathrm{Re} \gtrsim 2{,}800$.

In contrast to previous studies[28], in our more strongly curved pipe experiments, turbulent puffs could not be observed. The localized patches encountered, unlike puffs, do not have a persistent size but either shrink or expand. Figure 1a shows the fraction of the flow that is still turbulent at five different downstream locations as a function of the Reynolds number. These TF curves are S shaped and with downstream location, they shift to a higher Re. As illustrated by the flow visualization images (Fig. 1b), turbulence continues to decay with downstream distance. In particular, the TF curves become steeper, indicating that the coexistence regime is suppressed and that the transition becomes increasingly sharp. At the end of the 24,000-$R$-long pipe, coexistence only persists in a narrow Reynolds number range, $\Delta\mathrm{Re} \leq 40$, yet it is unclear if the flow has reached a statistical steady state or if with the downstream distance, this narrow regime may eventually vanish.

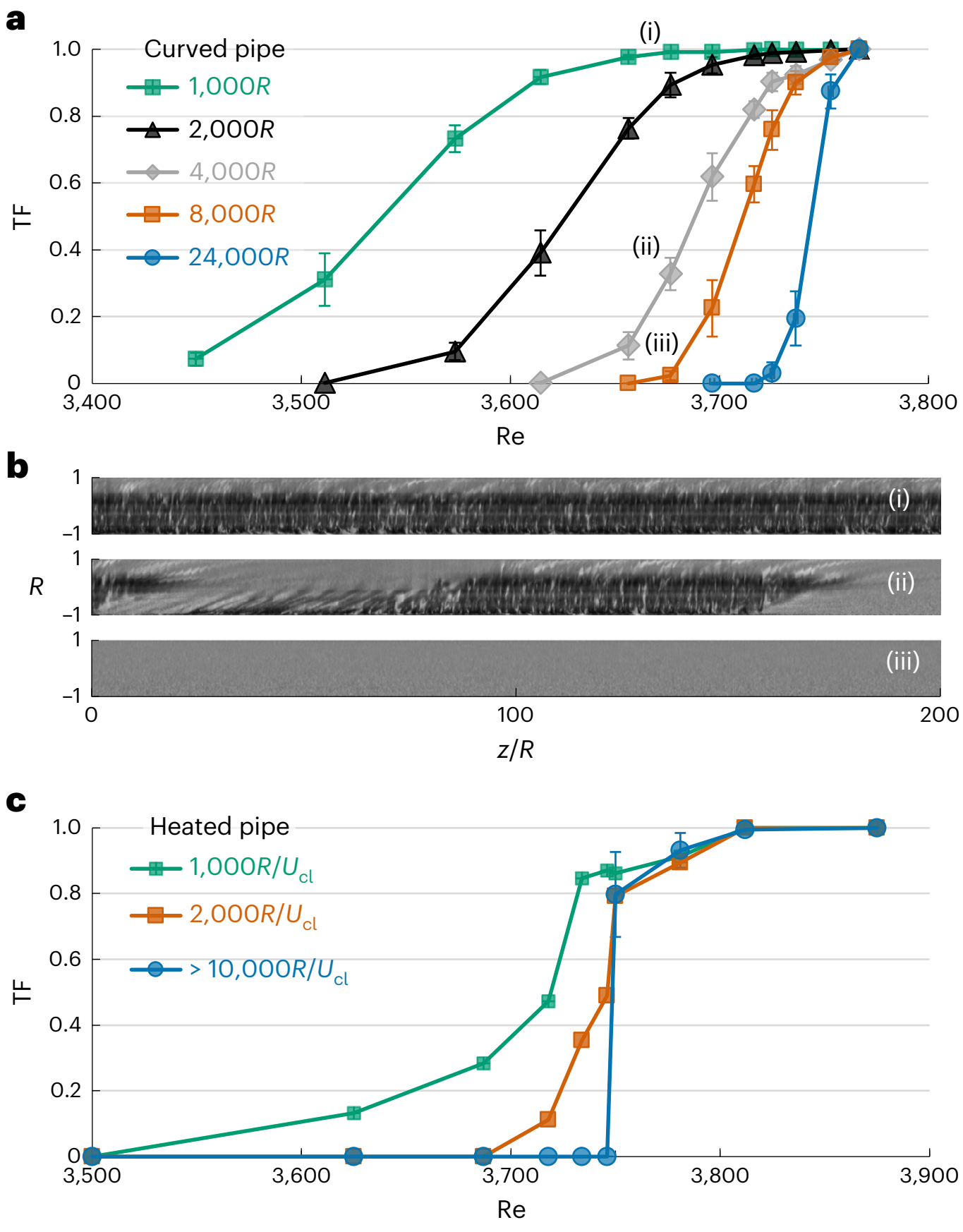

**Fig. 1 | Approach towards a discontinuous transition. a**, Reynolds number dependence of the TF in a curved (helical) pipe in experiments. Curves show the TFs measured at different downstream locations. The error bars indicate the standard deviation. The TF continues to adjust with downstream location and has not reached the statistical steady state at the end of the pipe (24,000$R$). The adjustment from laminar (TF = 0) to fully turbulent (TF = 1) flow occurs across a decreasing Re interval. **b**, An initially fully turbulent flow gradually laminarizes. The flow structure is visualized by adding small amounts of mica flakes (50 µm in size), turbulent regions show up in dark grey, whereas laminar ones appear bright. Turbulent transients persist over the first 8,000 pipe radii, until eventually the flow fully relaminarizes. **c**, TF as a function of Re, in a vertically heated pipe in direct numerical simulations. The error bars indicate the standard deviation. Here the TF initially adjusts and reaches a statistically steady state after approximately 10,000 advective time units. TFs below 0.8 cannot be sustained and the transition is, hence, discontinuous.

As shown next, the same qualitative trend is observed for the flow through heated pipes. In this case, we carry out the direct numerical simulations of a straight vertical pipe, 100$R$ in length and heated from the wall. The numerical parameters are listed in Extended Data Table 1, and further details regarding the code are provided in the Methods and ref. 30. The flow is driven by a pressure gradient in the upward direction, and at the same time, the fluid adjacent to the heated wall is of a lower density and experiences an additional upward buoyancy force. Like curvature (or electromagnetic forces), buoyancy (mixed convection) also delays the transition to turbulence, a circumstance that causes a sudden deterioration of heat transport[30–33] and mixing[34] in applications.

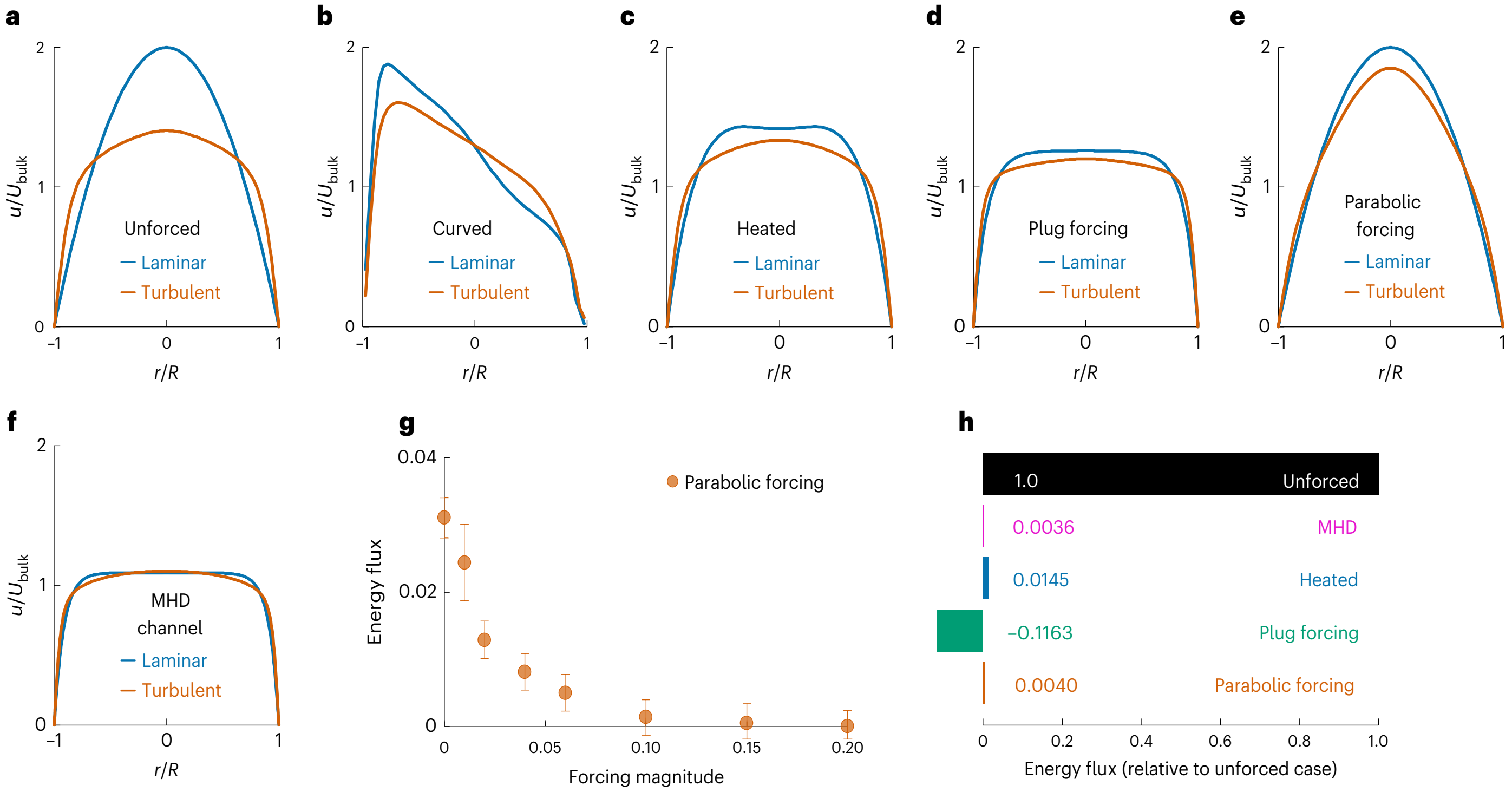

**Fig. 2 | Comparison of laminar and turbulent profiles. a–e**, (Unforced) pipe flow (**a**), curved pipe (**b**), heated pipe (**c**), plug forcing (**d**) and parabolic forcing (**e**). **f**, MHD channel flow. **g**, Energy advection across the upstream laminar–turbulent interface; an example of parabolic forcing is shown. The error bars indicate the standard deviation. With increasing body force amplitude, the energy advection goes to zero. **h**, Drop in energy advection across laminar–turbulent interfaces for different body forces compared with the unforced case.

In our simulations, we initialize the flow in a fully turbulent state and (like in the curved pipe case) monitor the TF as a function of time. Figure 1c shows the TF encountered at three different time instances as a function of Re. Also, in this case, the TF curves are initially S shaped and steepen as time proceeds, qualitatively resembling the phenomenology in curved pipes. In this case, however, owing to the periodic boundary conditions in the simulations, the observation times are not limited by the physical length of the pipe. At each Re, simulations were continued until eventually the data approached a statistical steady state (Fig. 1c, blue data points), often requiring more than $10^4$ advective time units ($R/U_{cl}$, where $U_{cl} = 2U$ is the laminar centreline velocity). For Re ≥ 3,750, turbulence persists with 80% of the domain or more being turbulent. Conversely, for Re < 3,750, although turbulence may persist for long times, the TFs eventually drop to zero, attesting that the transition is discontinuous. It is noteworthy that even in the case of heated pipes, puffs are absent.

The two body forces considered above affect the flow in entirely different ways. While curvature gives rise to a centrifugal force and distorts the profile radially, buoyancy accelerates the flow in the axial direction. Surprisingly, the impact on the transition to turbulence is qualitatively the same. To better understand why in both cases coexistence is suppressed and the transition becomes increasingly abrupt, we compare the respective laminar and turbulent velocity profile and recall that LTC arises from spatial coupling and requires energy exchange between laminar and turbulent regions.

As depicted in Fig. 2a, in the absence of body forces, the time-averaged turbulent velocity profile is plug like and has a far smaller peak velocity compared with the parabolic laminar flow. This large velocity difference facilitates spatial coupling and energy flux from laminar to turbulent regions. The situation is markedly different in the curved pipe case (Fig. 2b). The centrifugal force affects both laminar and turbulent flows, and overall, the effect of body force on the profile exceeds the profile deformation caused by turbulent eddies. Hence, the body force effectively reduces the difference between the laminar and turbulent profiles, and therefore, it limits the spatial energy transfer between the two states. As a consequence, LTC is suppressed. The transition is delayed until at a larger Re, the locally available shear suffices to sustain turbulence throughout the pipe irrespective of the spatial energy flux. Equally, in case of the heated pipe, buoyancy affects the shapes of both laminar and turbulent profiles. Again, the profile distortion caused by buoyancy far exceeds the distortion due to turbulence, and hence, also in this case (Fig. 2c), the laminar and turbulent profiles show much smaller differences than for ordinary pipe flow. In summary, the same argument applies: lacking spatial coupling and energy transfer, localized turbulent structures are suppressed, rendering the transition discontinuous.

To further investigate the general effect of body forces on the transition, we next consider two specifically designed cases. In both situations and in contrast to the previous two cases, here the laminar flow remains stable even for large force amplitudes, and hence, the transition remains subcritical throughout. The first forcing scheme has the tendency to make the profile more plug like, whereas the second tends to preserve a parabolic profile shape. The plug forcing, previously used to specifically target laminar–turbulent interfaces, is applied here globally. Unlike in ref. 18, it acts throughout, irrespective if the regions are laminar or turbulent. Its tendency to flatten the velocity profile is a property shared with the MHD pipe flow, where a transverse magnetic field induces a Lorentz force that decelerates the central flow and accelerates the near-wall fluid. Such plug-shaped velocity profiles are also typical for pipe flows of shear thinning fluids (Extended Data Fig. 2 shows the correspondence between the plug forcing profile and a shear thinning profile).

The parabolic forcing scheme has the opposite effect on turbulent flow and accelerates the flow at the pipe centre and tends to decelerate the near-wall fluid (Methods section and equation (9)). More precisely, the force is a linear damping term proportional to the local deviation from the parabolic profile, with a fixed damping coefficient $\alpha = 0.2$ (a coefficient of $\alpha = 0.04$ already has the same qualitative effect on transition). The force, hence, pushes the profile towards the laminar parabola, causing an acceleration in slower than in laminar regions and a deceleration in faster than in laminar regions. Out of the five body forces considered here, it is the only one that directly counters the deviation from a given profile. Visualizing the resulting velocity profiles, Fig. 2d,e shows that also the plug and parabolic forcing severely reduce the difference between the laminar and turbulent profile (compared with Fig. 2a).

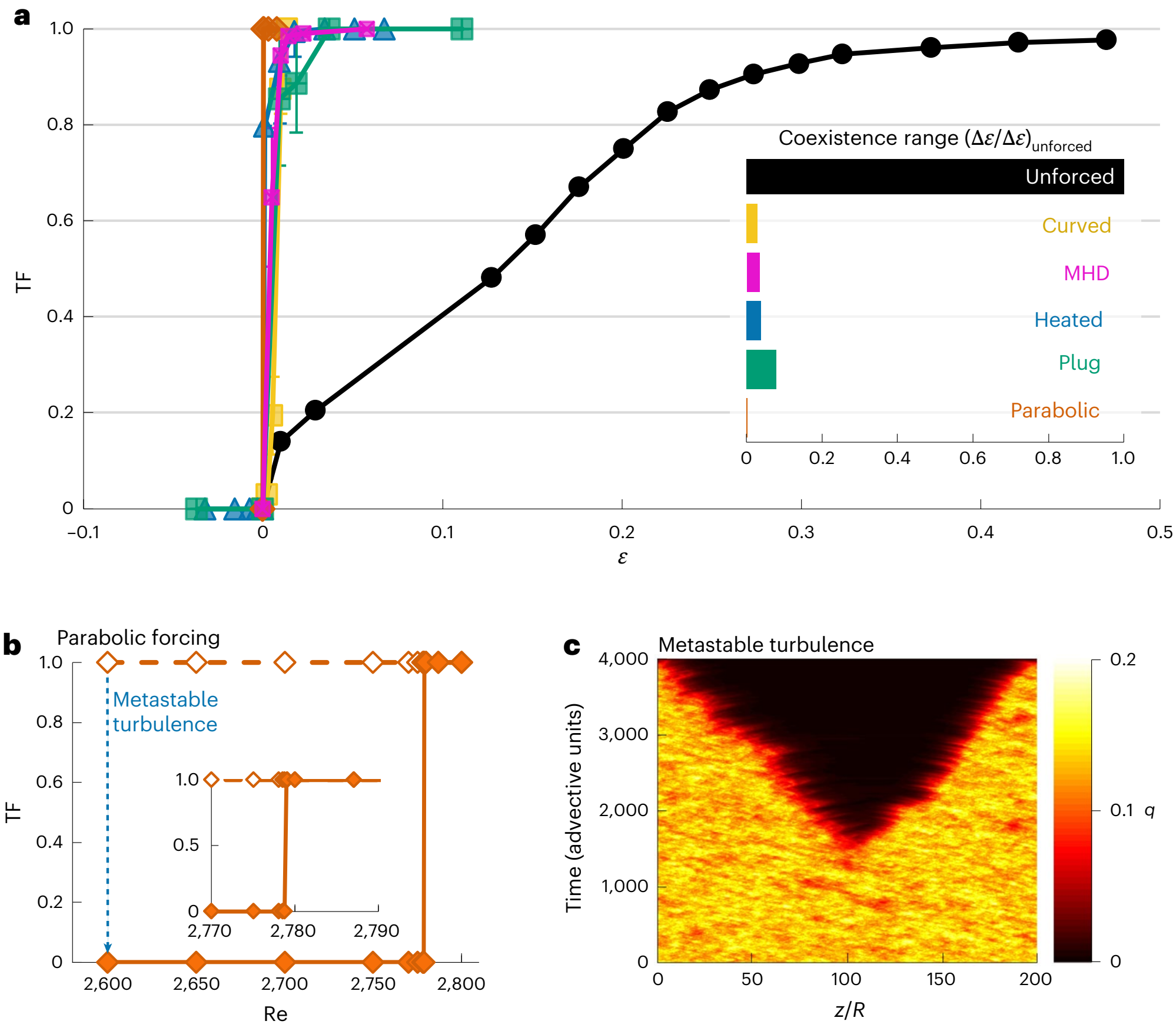

**Fig. 3 | Discontinuity, hysteresis and metastability. a**, TF as a function of the reduced Reynolds number $\epsilon$. The critical Reynolds number $Re_c$ is 2,040, 3,716.5, 7,090, 3,746, 6,750 and 2,778.75 for the unforced, curved, MHD, heated, plug and parabolic cases, respectively. The error bars indicate the standard deviation. The TFs for ordinary (unforced) pipe flow (black) are taken from experiments[39] and simulations[40] (via conversion of the reported friction factor data assuming the Blasius friction factor relation). Compared with this unforced case (black), the transition in the presence of the various body forces is sharp. The inset highlights the width of the LTC regimes (that is, the $\epsilon$ interval in which $0 < TF < 1$). **b**,**c**, For parabolic forcing, the transition is hysteretic (**b**). Full symbols show the TF at which the flow settles down, if initialized with TF = 0.5. Additionally, runs were carried out starting from fully turbulent flow (open symbols), which, in this case, persists even below the critical point. Here turbulence is metastable and after the nucleation of a laminar gap (**c**), the stable laminar state invades turbulence and the flow fully laminarizes.

To elucidate the consequences that this close match between the laminar and turbulent flow profiles has on the coexistence, we next compute the advection of kinetic energy across the laminar–turbulent interfaces. In our notation, a positive value corresponds to a net energy flux into the turbulent region (Methods; equation (6) provides the definition). As demonstrated for the parabolic force in Fig. 2g, this energy flux has its maximum value in ordinary (unforced) pipe flow and decreases monotonically with increasing forcing amplitude. As shown in Fig. 2h, this suppression of energy advection is a property that all investigated body forces have in common. In the heated pipe, the remaining advection amounts to barely 2% of the unforced level, and for the parabolic force, it is virtually reduced to zero. For plug forcing, energy advection has even become negative, showing that here kinetic energy is transported from turbulent regions to laminar ones.

Given the dependence of localized turbulent structures on energy advection from laminar regions, we would expect, just like for curved and heated pipes, that also in cases of the plug and parabolic forcing, puffs are absent and that generally the LTC regime is suppressed. To detect the transition threshold and the width of the coexistence regime, simulations were conducted for a range of Reynolds numbers, starting from a fully turbulent velocity field (obtained at a higher Re) and then continued until a statistical steady state was approached. Subsequently, the runs were repeated with a different initial condition, composed of laminar and turbulent regions (typically, TF ≈ 0.5) to detect hysteresis and metastability.

To compare this suppression and the abruptness of the transition for the various body forces, Fig. 3a shows TF as a function of the reduced Reynolds number, $\epsilon = (Re - Re_c)/Re_c$, where $Re_c$ denotes the respective critical point. Compared with unforced pipe flow, all four body forces display a sharp transition. While, for the former case, the width of the coexistence range, that is, the Re range for which $0 < TF < 1$, corresponds to $\Delta\epsilon \approx 0.5$, it only amounts to less than 0.04 for the body force cases. For the heated pipe and plug forcing (Extended Data Table 2), the TF jumps from about 0.8 directly to 0, attesting a discontinuous scaling.

We will next focus on parabolic forcing. In this case, the adjustment of the TF occurs within a Reynolds number interval of less than $\Delta Re = 0.25$ corresponding to $\epsilon = 0.0001$ or 0.01% of the critical value (Extended Data Table 3 lists the simulation parameters). The sharp switch between states and the discontinuous nature of this transition is further underlined by strong hysteresis (Fig. 3b). Coming from high Reynolds numbers, a fully turbulent flow persists far below the critical point (Fig. 3b, open diamonds). In these cases, the fully turbulent flow is found to be metastable. Once a laminar gap opens up (Fig. 3c shows an example for Re = 2,600), the laminar nucleus grows, and just like a seed crystal in a supercooled liquid, it completely eliminates the metastable phase. The transition, hence, shows all the characteristics expected at a discontinuous phase transition.

Finally, to demonstrate that the obtained results are not specific to pipe flow, we next test a different geometry, a channel flow subject to periodic boundary conditions and, hence, a flow unconfined in the planar directions. In this case, we study the effect of a Lorentz force. Specifically, we consider an electrically conducting fluid subject to a wall-normal magnetic field. The details of the computational scheme, domain size and resolution tests are provided in the Methods. In channel flow, like in pipes, the transition is subcritical and includes a broad LTC range starting from Re ≈ 1,500 down to Re ≈ 650, as discussed recently[13]. For simplicity and ease of comparison, here we use the same domain as in ref. 13. We intentionally pick a moderate forcing amplitude, selecting a Hartmann number (the ratio of the Lorentz force to the viscous force) of Ha = 12 (Methods; equation (15) provides the definition)—a value smaller than that in Earth's ionosphere or those typical for MHD flows of liquid metals or plasmas. As shown in Fig. 3a, the effect on transition is the same as for the other flows. Again, the coexistence regime is marginalized and the TF essentially jumps from a value close to one directly to laminar with decreasing Re.

Following the recent focus on the dominant role of LTC on the transition in basic shear flows and the analogy to directed percolation, the continuous nature of the transition may appear as the norm and the discontinuous route reported here as the exception. However, on a fundamental basis, a subcritical bifurcation is expected to map to a discontinuous phase transition[35,36]. In basic shear flows, this relation, however, is altered by LTC, resulting in the familiar continuous scenario. As our study shows, in the more complex setting of shear flows subjected to body forces, the theoretically expected correspondence between a subcritical bifurcation and a discontinuous phase transition is restored.

During the revision of our manuscript we have become aware of a study[37] proposing a possible mechanism for the discontinuities observed in the present work. On the basis of a simple one-variable model, which—in contrast to shear flows—is lacking (energy) advection and the puff (stripe) phenomenology, it is suggested that the level of turbulent fluctuations dictates if the transition is either continuous or discontinuous. It is important to note that in that study, the term 'turbulent fluctuations' differs from the common usage of the term and instead refers to spatial variations of the turbulence fluctuation level, that is, intermittency. The relevant question in this context is what drives and regulates spatiotemporal variations. As we have shown in the present study (Fig. 2), the key mechanism is spatial energy flux and, hence, a process that is omitted in the aforementioned model[37]. This point aside, the Couette simulations presented in the same study[37] demonstrate a suppression of turbulent stripes when a filtering of large-scale flows is applied. An analogous suppression of LTC has recently been reported for boundary layers in the presence of wall suction[38]. These examples suggest that beyond body forces, discontinuous shear flow transitions equally arise from other profile manipulations. The absence of strong laminar–turbulent fronts[37] indicates that here also a suppression of laminar–turbulent energy advection may occur.

## Online content

Any methods, additional references, Nature Portfolio reporting summaries, source data, extended data, supplementary information, acknowledgements, peer review information; details of author contributions and competing interests; and statements of data and code availability are available at https://doi.org/10.1038/s41567-025-03166-3.

## Methods

### Direct numerical simulations of vertical wall-heated pipe flow

In this section, we describe the mathematical model and numerical methods for conducting direct numerical simulations of vertical wall-heated pipe flow. The same numerical methods are adopted by the direct numerical simulations with plug and parabolic forcings.

We simulate an incompressible fluid in a circular pipe with a fixed mass flux. The governing equations of the fluid motion are the incompressible Navier–Stokes equations

$$\frac{\partial \mathbf{u}}{\partial t} + (\mathbf{u}\cdot\nabla)\mathbf{u} = -\nabla p + \frac{1}{\mathrm{Re}}\nabla^2\mathbf{u} + \mathbf{F}, \quad (1)$$

$$\nabla\cdot\mathbf{u} = 0, \quad (2)$$

where $\mathbf{u}$ is the velocity, $p$ is the pressure and $\mathbf{F}$ is the body forcing term applied to the fluid (in this case, the buoyancy force). The equations are non-dimensionalized by $R$ (pipe radius) and $2U$ (twice the bulk velocity) as the length and velocity scales, respectively. No-slip boundary conditions are imposed at the wall and the velocity field is assumed to be periodic in the axial direction.

The pipe is assumed to be vertical and heated from the wall. The flow is driven by an upward pressure gradient and the additional buoyancy force, whereas the mass flux is kept constant. The temperature of the wall $T_w$ is assumed to be fixed, whereas a uniform heat sink $\epsilon(t)$ is applied throughout the domain to maintain a fixed bulk temperature $T_b$, which is lower than $T_w$. Thus, the fluid near the wall is subject to an upward buoyancy force, resulting in a mean velocity profile flatter than the unforced profile. Under the Boussinesq approximation, the buoyancy forcing is given by (ref. 30 provides a detailed derivation)

$$\mathbf{F} = \frac{4C\Theta}{\mathrm{Re}}\hat{\mathbf{z}}, \quad (3)$$

where $\Theta = (T - T_c)/\Delta T$ is the non-dimensional temperature, $\Delta T = 2(T_w - T_b)$ and $T_c = T_w - \Delta T$ is a reference temperature. Parameter $C$ measures the ratio between the buoyancy force and the force that drives the laminar isothermal flow, and is defined as

$$C = \frac{\mathrm{Gr}}{16\mathrm{Re}} = \frac{\gamma g (T_w - T_b)R^2}{4U\nu}, \quad (4)$$

where $\mathrm{Gr} = 8\gamma g(T_w - T_b)R^3/\nu^2$ is the Grashof number, with $\gamma$ being the volume expansion coefficient and $g$ being the acceleration due to gravity. We use a constant value of $C = 3.5$ throughout this study.

The temperature field is governed by the following advection–diffusion equation:

$$\frac{\partial\Theta}{\partial t} + \mathbf{u}\cdot\nabla\Theta = \frac{1}{\mathrm{Re\,Pr}}\nabla^2\Theta - \frac{4}{\mathrm{Re\,Pr}}\frac{\epsilon(t)}{\epsilon_0}, \quad (5)$$

where $\mathrm{Pr} = \nu/\kappa$ is the Prandtl number ($\kappa$ is the thermal diffusivity and $\epsilon(t)$ is a heat sink that keeps the bulk temperature constant at $T_b$ and a laminar heating rate at $\epsilon_0 = 8\kappa(T_w - T_b)/R^2$). We use a constant $\mathrm{Pr} = 0.7$ as in ref. 30.

The open-source code openpipeflow[41] is used for the simulations. The code discretizes the equations in cylindrical coordinates $(r, \theta, z)$. Fourier–Galerkin expansion is used in the axial and azimuthal directions, and an eighth-order finite-difference method is used in the radial direction. The equations are integrated in time using a second-order predictor–corrector algorithm with a fixed time step of $1\times10^{-2}$ non-dimensional time units, determined by keeping the Courant number[42] below 0.5. A computational domain of $100R$ is used, which is enough to accommodate intermittent structures such as puffs and slugs. The simulation setups are summarized in Extended Data Table 1.

### Energy advection

LTC involves energy exchange between the laminar and turbulent regions. To quantify this energy flux into a turbulent slug, we consider the kinetic energy transport, defined as the cross-sectional integral of the kinetic energy times the axial velocity relative to the mean flow as

$$T(z) = \int\int \frac{1}{2}(u_r^2 + u_\theta^2 + u_z^2)(u_z - U)r\mathrm{d}r\mathrm{d}\theta, \quad (6)$$

at the upstream and downstream fronts of the slug. A similar definition has been used in ref. 18 to quantify the vorticity transport in a turbulent puff. The net energy flux into the slug is calculated as the difference in the energy transport at the two fronts:

$$\mathcal{T}_{\mathrm{net}} = \mathcal{T}_{\mathrm{up}} - \mathcal{T}_{\mathrm{down}}. \quad (7)$$

Mathematically, this is identical to integrating the advection term from the total kinetic energy budget equation within the slug. The upstream and downstream fronts are identified by setting a threshold value of the turbulent intensity following ref. 43. We use 10% of the mean turbulent intensity in the core region of the slug as the threshold, and we have verified that the results are insensitive to this threshold value.

### Plug forcing

The second body force—plug forcing—is designed to accelerate the flow near the wall and decelerate the flow near the pipe centre, whereas the mass flux is kept unchanged. This strategy has been used in previous studies for turbulent control[44]. Following ref. 44, the forcing is such that it forces a laminar velocity profile given by

$$u(\beta, r) = (1-\beta)\left[1 - \frac{\cosh(cr) - 1}{\cosh(c) - 1}\right], \quad (8)$$

where parameter $\beta$ represents the centreline velocity difference between the forced and unforced laminar profiles and $c$ is a constant to assure a constant mass flux. The body forcing $\mathbf{F}$ is then solved inversely given the target profile. Throughout this study, we use a constant value of $\beta = 0.37$.

The simulations are conducted using openpipeflow[41] with a fixed time step of $6\times10^{-3}$. The detailed simulations setups are summarized in Extended Data Table 2.

### Parabolic forcing

The third body force—parabolic forcing—is designed to have the opposite effect on the mean velocity profile compared with the buoyancy and plug forcings. It accelerates the flow near the pipe centre and decelerates the flow near the wall. This goal is achieved by choosing the following body force term:

$$\mathbf{F} = -\alpha(u_z - U_L)\hat{\mathbf{z}}, \quad (9)$$

where $u_z$ is the axial velocity component, $U_L = 1 - r^2$ is the laminar parabolic profile and $\alpha$ is the forcing magnitude. The parabolic forcing is actually a damping force that makes the mean velocity profile damp to a laminar parabola, whereas $1/\alpha$ represents the damping timescale. We fix $\alpha = 0.2$ throughout this study.

The simulations are conducted using the open-source code nsPipeFlow[45] with a fixed time step of $1\times10^{-2}$. The detailed simulation setups are summarized in Extended Data Table 3. We typically use a long domain of $200R$. However, as Re approaches the critical point, the relaxation time required to reach the statistically steady state becomes increasingly longer, and hence, the computational costs become larger. Nevertheless, to resolve the critical point, we decreased the domain size to $50R$. Note that in this case, no indications of puff-like structures are observed, and our simulations indicate that flows either settle to fully turbulent or fully laminar flow. For simulations of a fully turbulent flow, this shorter domain length of $50R$ can be considered large and more than sufficient to determine if the fully turbulent flow is sustained or recedes.

## MHD channel flow

An MHD channel flow is simulated using equation (1), where $\mathbf{F}$ is the Lorentz force in its dimensionless form and for a wall-normal ($\hat{\mathbf{y}}$) magnetic field, it is specified in equation (10). We are applying the customary quasi-static assumption[46,47], valid at a low magnetic Reynolds number and, hence, neglecting the induced magnetic fields. The equations are non-dimensionalized using $h$ (half-gap distance) and $3U_b/2$ as the length and velocity scales, respectively, with the Reynolds number defined as $\mathrm{Re} = \frac{3}{2}U_b h/\nu$. No-slip boundary conditions are imposed at the wall (direction $\hat{\mathbf{y}}$, with $y \in [-1, 1]$) as before and the velocity field is assumed to be periodic in two other orthogonal directions on the plane, with $\hat{\mathbf{x}}$ being the streamwise and $\hat{\mathbf{z}}$ being the spanwise directions. The streamwise mass flux is kept constant at $U_b = 2/3$. We use a version of openpipeflow[41] modified for planar flows, which we further modified to run with the Lorentz force ($\hat{\mathbf{y}}$), given by

$$\mathbf{F} = \frac{\mathrm{Ha}^2}{\mathrm{Re}}\mathbf{J} \times \hat{\mathbf{y}}, \tag{10}$$

where $\mathbf{J}$ is the induced current density, determined from the velocity field, the induced electric field (electric potential $V$) and the wall-normal magnetic field (here its magnitude has been absorbed into Ha as part of non-dimensionalization, to be defined later) via Ohm's law:

$$\mathbf{J} = -\nabla V + \mathbf{u} \times \hat{\mathbf{y}}. \tag{11}$$

Here $V$ is the electric potential subject to the insulating boundary conditions:

$$\frac{\partial V}{\partial y}(y = \pm 1) = 0. \tag{12}$$

This potential gradient is determined through the effective 'incompressibility' of the induced current density (charge conservation):

$$\nabla \cdot \mathbf{J} = 0, \tag{13}$$

which leads to the same computational problem as the incompressibility of the flow and the determination of the pressure gradient.

These equations follow from a set of approximations for an electrically conducting fluid at the limit of low magnetic Reynolds number[46,47]:

$$\mathrm{Re}_m = \bar{U} h \sigma \mu_0, \tag{14}$$

where $\bar{U}$ is the mean streamwise velocity, $\sigma$ is the electrical conductivity and $\mu_0$ is the magnetic permeability of the vacuum. Ha is defined as

$$\mathrm{Ha} = Bh\sqrt{\frac{\sigma}{\rho\nu}}, \tag{15}$$

where $B$ is the magnitude of the applied magnetic field and $\rho$ is the density of the fluid. Under this approximation valid at low $\mathrm{Re}_m$, the magnetic field is assumed to remain constant and induces a current on the fluid, and any magnetic field induced by the flow is ignored.

For this flow, the laminar state can be found analytically[48]:

$$U_{\mathrm{MHD}}(y) = \left(\frac{2}{3}\right)\frac{\mathrm{Ha}}{\mathrm{Ha} - \tanh \mathrm{Ha}}\left(1 - \frac{\cosh(\mathrm{Ha}y)}{\cosh \mathrm{Ha}}\right), \tag{16}$$

where we put the 2/3 prefactor for the bulk velocity to match with that of plane Poiseuille velocity.

To capture LTC in moderate computational domain sizes[49], the naturally assumed angle of turbulent stripes has to be taken into account and following recent studies[13,50]. Therefore, we tilted the rectangular computational domain by $\theta = 45°$ with respect to the streamwise direction. To be able to directly compare the MHD results to the LTC regime in the unforced channel flow case, which has been recently investigated[13], we simulate flows for the identical domain size, which, in units of $h$, corresponds to $10 \times 40$ ($L_{x'} \times L_{z'}$). The resolutions used for the MHD simulations are listed in Extended Data Table 4, and Extended Data Fig. 3 shows a comparison between the resolution we used for our studies and a resolution 50% higher in all directions of time- and space-averaged (in the homogeneous directions) velocity fluctuations as a function of distance to the wall. The equations are integrated in time with a second-order predictor–corrector scheme, with a variable time step such that the Courant–Friedrichs–Lewy condition[42] is satisfied with a Courant number of 0.5.

## Experimental setup

A schematic of the helical pipe experiment is shown in Extended Data Fig. 1. The helical pipe consisted of semiflexible polyurethane tubing with an inner diameter of $D = 2R = 8$ mm and a wall thickness of 2 mm, wound onto a metal tube of diameter $2A = 225$ mm. The coils were tightly wound such that the pitch of the helix was the same as the outer diameter of the tubing. As shown in Extended Data Fig. 1a, the resulting helical then had a radius ratio of $R/A = 0.0173$, a pitch of 12 mm (or 3R) and a length of $12{,}250D$. The working fluid was water, which was pushed through the pipe by a piston housed in a hollow cylinder, which was, in turn, driven by a precision linear actuator. Before entering the pipe, the water passed through a heat exchanger, which maintained the water temperature to 0.1 °C. The water temperature was measured before and after the helical section. The water viscosity was calculated using the mean temperature and standard temperature–viscosity tables. The set flow rate and calculated viscosity then sets $\mathrm{Re} = UD/nu$, which was accurate to 0.5%. Additionally, before entering the helical pipe, the water flows through four 10-cm-diameter loops followed by a straight section (length, 1 m). As increasing curvature delays the transition to turbulence, the flow exiting the loops and entering the straight section is laminar and, hence, remains laminar in the helical section. Near the entrance of the straight section, there is a small pin held in place with a magnetic actuator. When the pin is flush with the inner wall of the tube, the flow remains laminar. The actuator can be used to move the pin such that it is perpendicular to the wall, and transverse to the flow, which then causes the downstream flow to be turbulent (at the Re used in this study). This turbulent flow enters the helical section and its evolution can then be studied. The flow was visualized using coated mica platelets that align with the local shear and can be used to distinguish turbulent from laminar regions. The flow was then monitored in the helical pipe at five locations ($500D$, $1{,}000D$, $2{,}000D$, $4{,}000D$ and $12{,}000D$) downstream of the inlet, to determine the state of flow. This was done by illuminating a cross-section of the flow at each of these locations using a laser sheet (Extended Data Fig. 1b) and recording images using DSLR cameras at a rate of 60 Hz. The images were then analysed to extract the TF at these locations.

For each Re, the measurement was carried out as follows. First, a laminar flow was maintained and some images were acquired and averaged to obtain a background image of the laminar flow at that Re. Next, the pin was actuated to trigger turbulence. After waiting for a time required for turbulent structures to reach the last measurement position, recording was started at all locations and continued till the end of the piston stroke. At each location, each acquired image was subtracted from the background laminar flow. The standard deviation for each image was computed, and was close to zero for a laminar flow and a finite value for turbulent regions. Using a cut-off, the fraction of turbulent flow could be computed at each of these locations. This procedure is then repeated for different Re values. Additionally, to obtain the velocity profiles, the velocity vectors were measured in an equatorial plane of helical pipe using a particle image velocimetry system from LaVision.

## Data availability

Source data are available via Zenodo at https://doi.org/10.5281/zenodo.17514317 (ref. 51).

## Code availability

The numerical simulations were carried out using the open-source codes openpipeflow[41] and nsPipeflow[45].

## Acknowledgements

The work was supported by the Simons Foundation (grant number 662960, to B.H.).

## Author contributions

B.H. conceived the project. B.Y. carried out the numerical simulations of the flows subject to plug forcing, parabolic forcing and buoyancy. Y.Z. conducted the experiments. V.M. helped with the experimental design. G.Y. carried out the MHD channel simulations. E.M. supplied the heated pipe code and helped with the data interpretation. B.H. wrote the manuscript, with contributions from all authors.

## Funding

Open access funding provided by Institute of Science and Technology (IST Austria).

## Competing interests

The authors declare no competing interests.

## Additional information

**Extended data** is available for this paper at https://doi.org/10.1038/s41567-025-03166-3.

**Correspondence and requests for materials** should be addressed to Björn Hof.

**Peer review information** *Nature Physics* thanks Jeff Carpenter and the other, anonymous, reviewer(s) for their contribution to the peer review of this work.

**Reprints and permissions information** is available at www.nature.com/reprints.

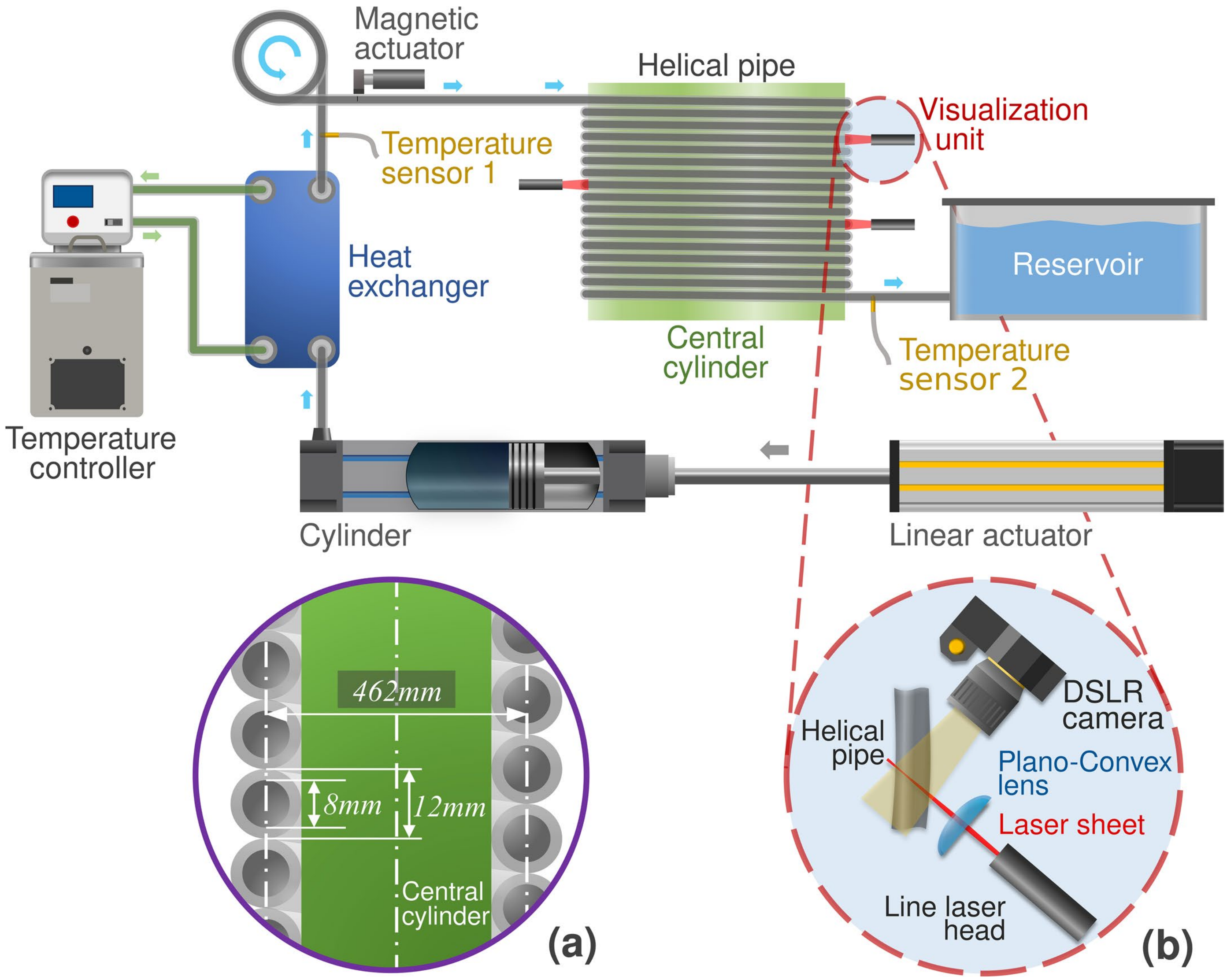

**Extended Data Fig. 1 | A schematic of the helical pipe setup.** Detail **a** shows a cross-section of the helical pipe and the relevant dimensions. Detail **b** shows how the flow was monitored at different locations along the length of the helical pipe.

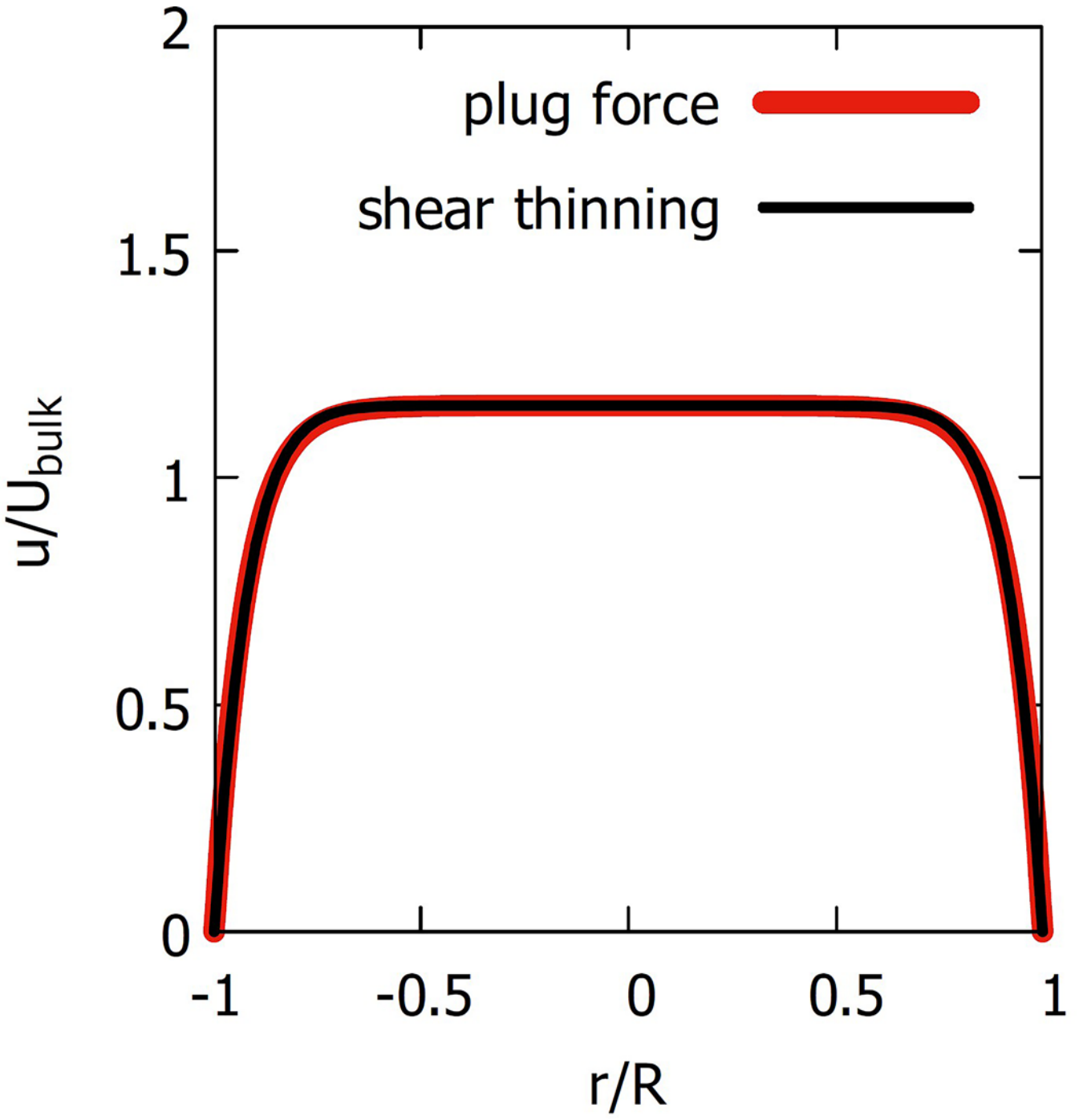

**Extended Data Fig. 2 | Comparison of the laminar profiles for a shear thinning fluid and for the plug force.** The shear thinning profile is obtained by assuming a power law fluid[52] with an exponent of $1/n = 12.5$.

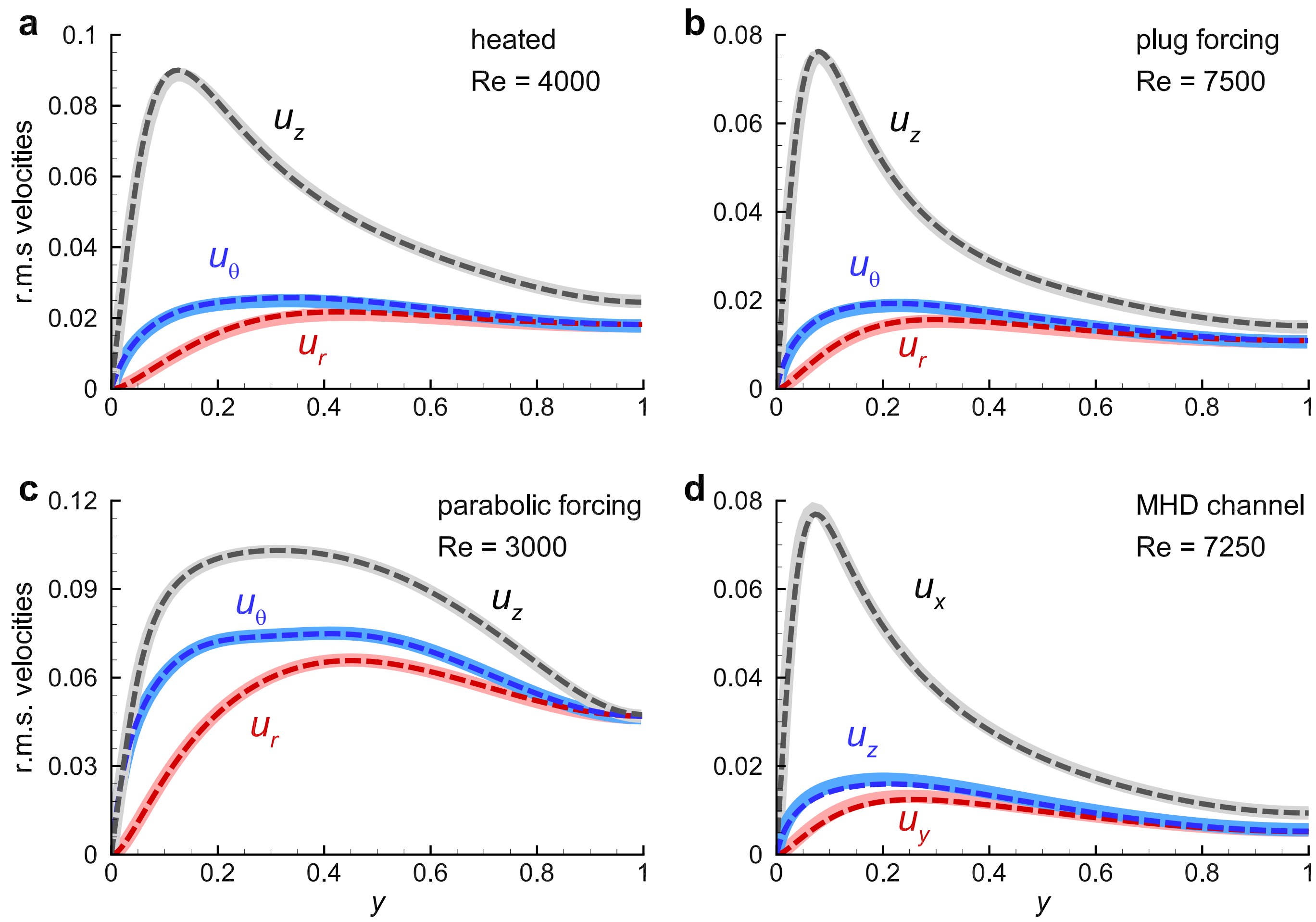

**Extended Data Fig. 3 | Grid resolution tests. a** heated pipe, **b** plug forcing, **c** parabolic forcing and **d** MHD channel at a Re above the critical point where the flow is fully turbulent. The grid convergence is checked by comparing the turbulent fluctuations (r.m.s. velocities) of our simulations and simulations with a resolution 50% higher in each direction. Solid lines with lighter colours are velocity fluctuations profiles from our simulations, while dashed lines with darker colours are from simulations with mesh refinement.

**Extended Data Table 1 | Heated pipe Reynolds numbers, domain size and resolutions**

| Re | $L_z(R)$ | $N$ | $K$ | $M$ | Re | $L_z(R)$ | $N$ | $K$ | $M$ |
|---|---|---|---|---|---|---|---|---|---|
| 3500 | 100 | 64 | 648 | 64 | 3746 | 100 | 64 | 648 | 64 |
| 3625 | 100 | 64 | 648 | 64 | 3750 | 100 | 64 | 648 | 64 |
| 3687 | 100 | 64 | 648 | 64 | 3781 | 100 | 64 | 648 | 64 |
| 3718 | 100 | 64 | 648 | 64 | 3812 | 100 | 64 | 648 | 64 |
| 3734 | 100 | 64 | 648 | 64 | 3875 | 100 | 64 | 648 | 64 |
| 3742 | 200 | 64 | 1280 | 64 | 4000 | 100 | 64 | 648 | 64 |

From left to right: Reynolds number, domain size, number of radial grid points, number of non-negative axial Fourier modes and number of non-negative azimuthal Fourier modes. Note that in physical space, there are 3*K* and 3*M* grid points in axial and azimuthal directions, respectively.

**Extended Data Table 2 | Plug forcing Reynolds numbers, domain size and resolutions**

| Re | $L_z(R)$ | $N$ | $K$ | $M$ | Re | $L_z(R)$ | $N$ | $K$ | $M$ |
|---|---|---|---|---|---|---|---|---|---|
| 6500 | 100 | 96 | 1024 | 96 | 6937 | 100 | 96 | 1024 | 96 |
| 6750 | 100 | 96 | 1024 | 96 | 7000 | 100 | 96 | 1024 | 96 |
| 6812 | 100 | 96 | 1024 | 96 | 7250 | 100 | 96 | 1024 | 96 |
| 6865 | 100 | 96 | 1024 | 96 | 7500 | 100 | 96 | 1024 | 96 |

From left to right: Reynolds number, domain size, number of radial grid points, number of non-negative axial Fourier modes and number of non-negative azimuthal Fourier modes. Note that in physical space, there are $3K$ and $3M$ grid points in axial and azimuthal directions, respectively.

**Extended Data Table 3 | Parabolic forcing Reynolds numbers, domain size and resolutions**

| Re | $L_z(R)$ | $N$ | $K$ | $M$ | Re | $L_z(R)$ | $N$ | $K$ | $M$ |
|---|---|---|---|---|---|---|---|---|---|
| 2500 | 200 | 64 | 1536 | 48 | 2778 | 50 | 64 | 384 | 48 |
| 2550 | 200 | 64 | 1536 | 48 | 2778.5 | 50 | 64 | 384 | 48 |
| 2575 | 200 | 64 | 1536 | 48 | 2778.75 | 50 | 64 | 384 | 48 |
| 2600 | 200 | 64 | 1536 | 48 | 2779 | 50 | 64 | 384 | 48 |
| 2650 | 200 | 64 | 1536 | 48 | 2780 | 50 | 64 | 384 | 48 |
| 2700 | 200 | 64 | 1536 | 48 | 2781 | 50 | 64 | 384 | 48 |
| 2750 | 50 | 64 | 384 | 48 | 2787 | 50 | 64 | 384 | 48 |
| 2775 | 50 | 64 | 384 | 48 | 2800 | 200 | 64 | 1536 | 48 |
| 2776 | 50 | 64 | 384 | 48 | 3000 | 200 | 64 | 1536 | 48 |

From left to right: Reynolds number, domain size, number of radial grid points, number of non-negative axial Fourier modes and number of non-negative azimuthal Fourier modes. Note that in physical space, there are $3K$ and $3M$ grid points in axial and azimuthal directions, respectively.

**Extended Data Table 4 | Magnetohydrodynamic channel flow, Hartmann numbers and resolutions**

| Ha | $N$ | $K$ | $M$ |
|---|---|---|---|
| 0 | 65 | 48 | 162 |
| 12 | 135 | 128 | 640 |

From left to right: Hartmann number, number of wall-normal grid points, number of non-negative Fourier modes in the short direction and the number of non-negative Fourier modes in the long direction. Note that in physical space, there are 3$K$ and 3$M$ grid points in the short and long directions, respectively. Reynolds numbers varied between Re = 7090 to 7500 for the magnetohydrodynamic (Ha = 12) runs. An unforced (Ha = 0) simulation at Re = 700 was used as the reference for the unforced energy flux.